\documentclass[twocolumn,showpacs,preprintnumbers,amsmath,amssymb,prl]{revtex4}
\usepackage{amssymb}
\usepackage{stmaryrd}
\usepackage{txfonts}
\usepackage{amsmath}
\usepackage{CJK}

\usepackage{graphicx}
\usepackage{epstopdf}
\usepackage{dcolumn}
\usepackage{bm}
\usepackage{ulem}

\begin{document}
\begin{CJK*}{GB}{SongMT}
\CJKfamily{gbsn}

\preprint{}

\title{Effective spontaneous $\mathcal{PT}$-symmetry breaking in hybridized metamaterials}

\author{Ming Kang$^{1}$, Fu Liu$^{1}$, Jensen  Li$^{2}$}

\email{j.li@bham.ac.uk}

\affiliation{$^1$Department of Physics and Materials Science, City University of Hong Kong, Tat Chee Avenue, Kowloon Tong, Hong Kong, China\\
$^2$School of Physics and Astronomy, University of Birmingham, Birmingham B15 2TT, UK}

\date{\today}

\begin{abstract}

We show that metamaterials can be used as a testing ground to investigate spontaneous symmetry breaking associated with non-Hermitian quantum systems. By exploring the interplay between near-field dipolar coupling and material absorption or gain, we demonstrate various spontaneous breaking processes of the  $\mathcal{PT}$-symmetry for a series of effective Hamiltonians associated to the scattering matrix. By tuning the coupling parameter, coherent perfect absorption, laser action and gain-induced complete reflection ($\pi$ reflector) by using an ultra-thin metamaterial can be obtained. Moreover, an ideal $\mathcal{PT}$-symmetry can be established effectively in a passive system by exploring the balance between scattering and absorption loss using metamaterials.
\end{abstract}


\maketitle
\end{CJK*}
Metamaterials, periodic arrangements of artificial subwavelength atoms, have paved the way for realizing different novel concepts and applications in optics, such as negative refraction, sub-diffraction imaging and invisibility cloaking\cite{R01,R02,R03,R04}. Instead of using the macroscopic material parameters for description, many of the optical properties, including metamaterial-based chirality\cite{R05,R06} and electromagnetically induced transparency (EIT)\cite{R07,R08,R09,R10,R11,R12,R13,R14}, can be understood by studying the coupling between different magnetic and electric artificial atoms in the so called hybridization picture \cite{R07,R12,R15,R16}. Similar to optical lattices with controllable coupling\cite{R17,R18,R19}, the design flexibility in the coupling between artificial atoms opens up a new perspective to realize quantum simulators to achieve effects which are difficult to obtain in quantum systems\cite{R17,R18}. Here, we theoretically investigate the possibility in constructing a metamaterial system with an effective $\mathcal{PT}$-symmetric Hamiltonian which can still show real eigenenergies. One of the most interesting properties of a  $\mathcal{PT}$-symmetric Hamilton is the existence of a phase transition arising from spontaneous symmetry breaking  by tuning an external parameter, defined as the coalescence of two eigenvalues and their corresponding eigenvectors\cite{R20}. Unlike the previously considered symmetry breaking effects of metamaterials\cite{R21}, such a spontaneous symmetry breaking is implicit in nature. It means that the phase transition occurs without structurally destroying a specific symmetry (e.g. mirror, rotational symmetry). The external parameter being tuned still respect the $\mathcal{PT}$-symmetry.

During the past few years, $\mathcal{PT}$-symmetric Hamiltonians have in fact been extensively investigated within the framework of optics. The most direct analog is  represented by a pair of optical waveguides with a balanced gain and loss profile\cite{R19,R22,R23,R24,R25,R26,R27,R28,R29,R30,R31}. Spontaneous $\mathcal{PT}$-symmetry breaking has also been theoretically and experimentally investigated\cite{R19,R22,R23,R24,R25,R26,R27,R28,R29,R30,R31}. Intriguing phenomena, such as asymmetric transmission\cite{R19,R24}, power oscillation \cite{R24}, unidirectional invisibility\cite{R19,R29}, coherent perfect absorption and laser\cite{R30} have been predicted. Moreover, extending from an ideal $\mathcal{PT}$-symmetric system with balanced gain and loss, non-ideal $\mathcal{PT}$-phase transition of a passive system has also been proposed and experimentally demonstrated \cite{R25,R27}. Unconventional loss-induced optical transparency is demonstrated  around the phase transition point\cite{R25}. Actually, $\mathcal{PT}$-symmetry has not been discussed within the context of metamaterials. By employing metamaterials, the larger flexibility in designing the coupling between different artificial atoms gives us additional possibilities in realizing non-trivial $\mathcal{PT}$-symmetry. Here, we show that by mapping a metamaterial consisting of a bright electric and a dark magnetic atom to an effective Hamiltonian, we can establish an ideal spontaneous symmetry breaking transiting from little absorption to complete absorption even in a passive system. It relies on the inverse relationship that scattering is much larger than loss for a bright electric atom while the reverse is true for a dark atom. Such a system has been previously considered to achieve EIT. However, we are interested in the phase transition to obtain complete absorption\cite{R32,R33} across a threshold in the coupling parameter, in contrast to the high transmission for EIT emerging from the strong coupling regime\cite{R07}. Moreover, by incorporating a small material gain into the dark atom of the ultrathin metamaterial, we can obtain laser action\cite{R34,R35,R36}  and gain-assisted complete reflection by considering the spontaneous symmetry breaking effect of a class of Hamiltonians associated with the scattering matrix.

The discussion of $\mathcal{PT}$-symmetric Hamiltonians originates from Bender and colleague's work, showing that a Hamiltonian $\mathbf{H}$ respecting a parity-time symmetry, \textit{i.e.} $[\mathcal{PT},\mathbf{H}]=0$, can have completely real eigenenergies\cite{R20}. To find the analog of  $\mathcal{PT}$-symmetry in a metamaterial system, the simplest way is to look for a pair of coupled  artificial atoms, since their response can be understood within a hybridization picture in quantum mechanics as a two state problem, and the $\mathcal{PT}$-symmetry in two-state quantum system is already well studied\cite{R20}. Here, we consider a metamaterial constituting a planar array of atoms. Each unit cell consists of an electric atom coupled to a magnetic atom through the near fields. For simplicity, we assume normal incidence and there is no polarization conversion for the incoming plane waves. The optical response of such a metamaterial can be expressed using the scattering matrix $\mathbf{S}$ as

\begin{equation}
\left(\begin{array}{cc}
    b_{+}  \\
    b_{-} \\
  \end{array}
\right)=\mathbf{S}\left(\begin{array}{cc}
    a_{+}  \\
    a_{-} \\
  \end{array}
\right)=\left(\begin{array}{cc}
    t & r \\
    r & t \\
  \end{array}
\right)\left(\begin{array}{cc}
    a_{+}  \\
    a_{-} \\
  \end{array}
\right)
\end{equation}
\noindent where $a_{\pm}$ and $b_{\pm}$ are the E-field amplitudes of the incoming and outgoing waves in the positive and negative directions. The same scattering process has a microscopic representation using the atomic responses and their associated far-field radiations. By further assuming each atom is dominated by only one dipolar resonating mode, the atomic responses can be summarized as
\begin{equation}
(\mathbf{H_{0}}-\omega\mathbf{I})\left( \begin{array}{cc}
    \tilde{p}\\
     i\tilde{m}\\
  \end{array}
\right)=\left( \begin{array}{cc}
    \sqrt{\gamma_{1}^{s}}(a_{+}+a_{-})\\
    i\sqrt{\gamma_{2}^{s}}(a_{+}-a_{-})\\
  \end{array}
\right)
\end{equation}
\noindent while the outgoing fields and the atomic responses are related to each other by
\begin{equation}
(\mathbf{H_{2}}-\omega\mathbf{I})\left( \begin{array}{cc}
    \tilde{p}\\
     i\tilde{m}\\
  \end{array}
\right)=\left( \begin{array}{cc}
    \sqrt{\gamma_{1}^{s}}(b_{+}+b_{-})\\
     i\sqrt{\gamma_{2}^{s}}(b_{+}-b_{-})\\
  \end{array}
\right)
\end{equation}
\noindent where
\begin{equation}
\mathbf{H_{n}}=\left(\begin{array}{cc}
    \omega_{1}-i\gamma_{1}  & \kappa \\
    \kappa  & \omega_{2}-i\gamma_{2}\\
  \end{array}
\right)+n\left(\begin{array}{cc}
    i\gamma_{1}^{s}  & 0 \\
    0  & i\gamma_{2}^{s}\\
  \end{array}
\right)
\end{equation}
\noindent In the above equations, $\gamma_{i}=\gamma_{i}^{s}+\gamma_{i}^{loss}$($i=1,2$) represents the total resonating linewidth for each atom where $\gamma_{i}^{s}$, a positive number, represents the part of scattering and $\gamma_{i}^{loss}$ is the part due to absorption loss when it is positive.
We also normalize the electric dipole $\tilde{p}$ and the magnetic dipole $\tilde{m}$ by the oscillator strengths (with details in establishing the above model given in Appendix) and assume the two resonant frequencies at the same frequency $\omega_{1}=\omega_{2}=\omega_{0}$. In fact, Eq. (4) defines a series of Hamiltonian $\mathbf{H_{n}}$ with meaning in the following. By taking an appropriate arithmetic average of Eq.(2) and Eq.(3), one can prove that $\mathrm{det}(\mathbf{H_{n}}-\omega\mathbf{I})=0$ yields  an eigenvalue $(n-2)/n$  of matrix $\mathbf{S}$. We can also obtain
\begin{equation}
\mathrm{det}(\mathbf{S})=\frac{\mathrm{det}(\mathbf{H_{2}}-\omega\mathbf{I})}{\mathrm{det}(\mathbf{H_{0}}-\omega\mathbf{I})}=2\frac{\mathrm{det}(\mathbf{H_{1}}-\omega\mathbf{I})}{\mathrm{det}(\mathbf{H_{0}}-\omega\mathbf{I})}-1
\end{equation}
\noindent For example, $\mathrm{det}(\mathbf{H_{0}}-\omega\mathbf{I})=0$ leads to the pole of $\mathbf{S}$, which corresponds to laser action\cite{R30,R34,R35,R36}. $\mathrm{det}(\mathbf{H_{2}}-\omega\mathbf{I})=0$ implies a zero of $\mathbf{S}$ to completely absorb the incoming EM waves, namely coherent perfect absorption (CPA)\cite{R30,R32,R33}. For a metamaterial with mirror symmetry in the propagation direction, the two eigenvectors of $\mathbf{S}$ are $(a_{+},a_{-})=(1,1)$ with eigenvalue $t+r$ and $(1,-1)$ with eigenvalue $t-r$ by defining $t/r$ as the complex transmission/reflection coefficient for electric field with respect to the mirror plane. If we further confine our discussion to the coupling between a bright electric and a dark magnetic atom, we have the additional requirement $t=1+r$. Then, $\mathrm{det}(\mathbf{H_{1}}-\omega\mathbf{I})=0$ implies a complete reflection $r=-1$ and $t=0$. In general, a particular $n$/eigenvalue of $\mathbf{S}$ denotes a particular optical phenomenon and it corresponds to solving the eigenvalue problem of $\mathbf{H_{n}}$. If the eigenvalue of $\mathbf{H_{n}}$ is real, the phenomenon can occur at a real frequency with a pole/zero on the corresponding optical spectrum.  Therefore, by considering the spontaneous symmetry breaking process of a $\mathcal{PT}$-symmetric $\mathbf{H_{n}}$, one can observe the phase transition when the real eigenfrequencies are turned into complex conjugates or vice versa. For our $2\times2$ Hamiltonian, the P-operation switches rows and columns of the matrix while the T-operation takes complex conjugate on the elements. A $\mathcal{PT}$-symmetric $\mathbf{H_{n}}$ is defined by $[\mathcal{PT},\mathbf{H_{n}}]=0$. We have to emphasize that $\mathcal{PT}$-symmetry being discussed here is an effective one through the establishment of an effective $\mathcal{PT}$-symmetric Hamiltonian, which is different to the approach employing complex potentials\cite{R20}.

As an example, a $\mathcal{PT}$-symmetric $\mathbf{H_{2}}$ corresponds to the condition $\gamma_{2}^{loss}-\gamma_{2}^{s}=\gamma_{1}^{s}-\gamma_{1}^{loss}$, which can be satisfied by a representative hybridized metamaterial with a bright electric atom (scatters most) and a dark magnetic atom(absorbs most). We should emphasize this kind of balance between scattering loss and absorption loss to obtain effective $\mathcal{PT}$-symmetric Hamiltonian is new and different to previous approaches with balanced gain and loss\cite{R19,R22,R23,R24,R25,R26,R27,R28,R29,R30,R31}. For a fixed set of parameters, \textit{e.g.} $\gamma_{1}=15$ THz, $\gamma_{1}^{loss}=1.75$ THz ($\gamma_{1}^{s}=13.25$ THz), $\gamma_{2}\approx\gamma_{2}^{loss}$, this condition is represented by the vertical red line in Fig. 1 at $\gamma_{2}=11.5$ THz to indicate a $\mathcal{PT}$-symmetric $\mathbf{H_{2}}$. Then, the spontaneous symmetry breaking process can be observed by changing the coupling coefficient $\kappa$ (still respecting the $\mathcal{PT}$-symmetry) across a threshold value. The eigenfrequencies of $\mathcal{PT}$-symmetric $\mathbf{H_{2}}$ are real(solid red line) when $\kappa>\gamma_{2}$, and becomes a complex conjugate pair(dotted red line) when $\kappa<\gamma_{2}$. The splitting of the real/imaginary part of the eigenfrequencies with varying $\kappa$ is showed in the inset. This behavior reveals itself through the absorption coefficient when the metamaterial is excited by the incident waves of $(a_{+},a_{-})=(1,1)$, giving CPA for $\kappa>\gamma_{2}$ at the two real frequencies. We note that such a configuration of coupling between a bright and a dark atom (without the $\mathcal{PT}$-symmetric condition satisfied in general) has been used to give EIT in the strong coupling regime\cite{R07}. We endow the same system to visualize effective $\mathcal{PT}$-symmetry breaking and focus on its optical properties in the regime near the critical point.
\begin{figure}
\centerline{
\includegraphics[width=8.0cm]{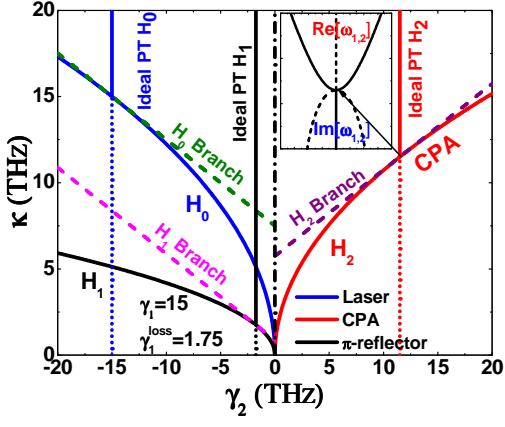}}
\caption{(Color online) Parametric evolution of the metamaterial for different $\mathbf{H_{n}}(n=0,1,2)$; Inset: Splitting of eigenvalues for $\mathbf{H_{2}}$ with varying $\kappa$. Vertical solid lines indicate the real eigenvalues case for ideal $\mathcal{PT}$-symmetric $\mathbf{H_{n}}$ above the phase transition threshold, short dotted lines illustrate the complex conjugate pair eigenvalues case for ideal $\mathcal{PT}$-symmetric $\mathbf{H_{n}}$ below the phase transition threshold. Inset gives the eigenvalues with varying $\kappa$ for $\mathcal{PT}$-symmetric $\mathbf{H_{2}}$. Curved solid lines show the real eigenvalue condition indicating the optical properties for non-ideal $\mathcal{PT}$-symmetric $\mathbf{H_{n}}$, short dashed lines show branching condition for non-ideal $\mathcal{PT}$-symmetric $\mathbf{H_{n}}$.}
\end{figure}
\noindent In the following, we demonstrate the interesting optical properties associated with  $\mathcal{PT}$-symmetry breaking with metamaterial structures. In particular, undergoing the phase transition, the metamaterial becomes perfectly absorptive with real frequencies. The bright atom (inset of Fig. 2(a)) is a silver strip of length $L1=150$ nm, width $W1=50$ nm and thickness $t=20$ nm. Fig. 2 (a) shows the electric field near one end of the strip with numerical simulation (symbols) and theoretical model (lines). The numerical simulation is carried out using finite element analysis (Comsol MultiPhysics) with Drude model of silver being $\epsilon=\epsilon_{\infty}-\omega_{p}^2(\omega^{2}+i\gamma_{p}\omega)^{-1}$, where $\omega_{p}=2.1961\times10^{15}$ Hz, $\gamma_{p}=4.3439\times10^{12}$ Hz, and $\epsilon_{\infty}=3.7$. The theoretical result is fitted using a Lorentz model $E_{s}\propto(\omega_{0}-\omega-i\gamma_{1})^{-1}(\gamma_{1}-\gamma_{1}^{loss})E_{0}$, where $\omega_{0}=402.5$ THz, $\gamma_{1}=15$ THz, and $\gamma_{1}^{loss}=1.75$ THz,
$E_{s(0)}$ is the scattering(incident) $\mathbf{E}$ field. $\gamma_{1}^{loss}$ can be estimated by the probing the response of the bright atom with and without metal loss. The dark atom is a magnetic resonator tuned to the same resonating frequency with two parallel silver strips and a dielectric gap material ($\epsilon=2.25+i\epsilon^{\prime\prime}$  with $\epsilon^{\prime\prime}$ introduced to represent a slight absorption or gain) in between. The geometric parameters are $L2=100$ nm, $W2=30$ nm, $t=20$ nm, as showed in the inset of Fig. 2(b). Due to the large quality factor of the dark mode resonance, a small $\epsilon^{\prime\prime}$ of the gap material can effectively alter the half-linewidth $\gamma_{2}$. As shown in Fig. 2 (b), numerically calculated   field (symbols) near one of the metal strip ends is again fitted well with theoretical Lorentz model (solid line) for $\epsilon^{\prime\prime}=0,-0.1$ (hollow, solid symbols). Comparing between numerical and analytical results with different $\epsilon^{\prime\prime}$, a simple approximated formula between $\gamma_{2}$
and $\epsilon^{\prime\prime}$ can be established as $\gamma_{2}=1.75+22.5\epsilon^{\prime\prime}$. We need to emphasize that, dark atom at magnetic resonance can provide flexibility to realize $\gamma_{2}^{loss}>>\gamma_{2}^{s}$, bright atom provide flexibility to realize $\gamma_{1}^{s}>\gamma_{1}^{loss}$, and hybridized system with these two atoms can provide a opportunity to satisfy the requirement of $\mathcal{PT}$-symmetric condition for $\mathbf{H_{2}}$. Now, we hybridize these two atoms within a metamaterial with a square lattice of periodicity $d=400$ nm and a distance $h$ in separating the two atoms, as shown in Fig. 2(c). The change of distance $h$ between the two meta-atoms can tune the coupling $\kappa$ between them, but there exists no simple relationship between them, as shown in Fig. 2(d). The $\kappa$ can be calculated from the   bright meta-atom probe response for $\epsilon^{\prime\prime}=0$ by changing the distance $h$ between them. The fitted formula can represent main variation character of $\kappa$, but exhibits slight deviation in large $\kappa$ region.
\begin{figure}
\centerline{
\includegraphics[width=8.0
cm]{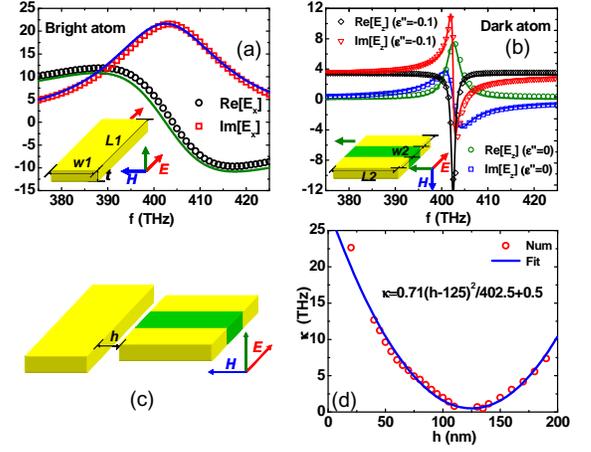}}
\caption{(Color online) Analytical (solid line) and numerical (scatter line) spectral response of the E-field probe for bright(a) and dark(b) meta-atom to establish the atomic Lorentz model with fitted parameters ($\omega_{0}$, $\gamma_{1}$, $\gamma_{1}^{loss}$, and $\gamma_{2}$). (c) and (d): The hybridized metamaterial under consideration and the relationship between the coupling coefficient $\kappa$ and the distance $h$ between the two meta-atoms.}
\end{figure}

\begin{figure}
\centerline{\includegraphics[width=8.0 cm]{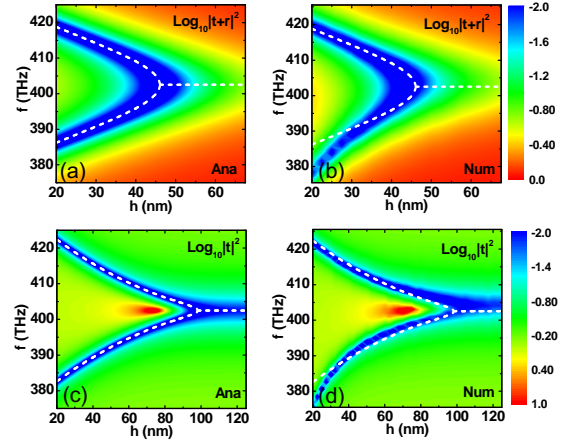}} \caption{(Color
online) Observation of effective spontaneous $\mathcal{PT}$-symmetry breaking for $\mathbf{H_{2}}$ and $\mathbf{H_{1}}$. Analytical model (with fitted parameters from Fig. 2) (a) and numerical(b) outgoing $|t+r|^2$ spectra versus frequency and coupling distance. Analytical(c) and numerical(d) transmission spectra versus frequency and coupling distance.}
\end{figure}

Now, we illustrate the spontaneous symmetry breaking with fixed geometry of the meta-atoms with constant $\omega_{0}$, $\gamma_{1}$, and $\gamma_{1}^{loss}$. The variable parameters are the coupling coefficient $\kappa$ (or equivalently $h$) between the two atoms and $\gamma_{2}$ (through the tuning of $\epsilon^{\prime\prime}$). We can then choose a particular $\gamma_{2}$ to satisfy $\mathcal{PT}$-symmetry requirement of $\mathbf{H_{n}}$. For a $\mathcal{PT}$-symmetric $\mathbf{H_{2}}$, we need $\gamma_{2}=2\gamma_{1}^{s}-\gamma_{1}=11.5$ THz, \textit{i.e.} $\epsilon^{\prime\prime}=0.433$, which is a passive configuration to display the effective $\mathcal{PT}$ symmetry. In the light of the above theoretical discussion, the eigenvalue problem for a $\mathcal{PT}$-symmetric $\mathbf{H_{2}}$ is equivalent to the zeros of the $\mathbf{S}$, which can be easily observed in the coherent outgoing $|r+t|^2$ spectra when we change the parameter $h$. The $\mathcal{PT}$-phase transition point in this case occurs at $\kappa_{PT}=11.5$ THz ($h=46.0$ nm). Below the critical point, $\kappa_{PT}<11.5$ THz ($h>46.0$ nm), the eigenvalues of $\mathbf{H_{2}}$ is a complex conjugate pair, there are no crossing in real frequency axis, so we cannot observe CPA in this region, shown in Fig. 3(a). As we increase (decrease) $\kappa(h)$, at $\kappa_{PT}=11.5$ THz($h=46.0$ nm), the two eigenvalues of $\mathbf{H_{2}}$ collapse to one point in the real frequency axis, \textit{i.e.} CPA($|r+t|=0$) emerges in the spectra. When $\kappa$ passes over the $\kappa_{PT}$, there are two real frequencies satisfies the eigenvalue solution of $\mathbf{H_{2}}$, CPA branching phenomenon can be clearly shown in Fig. 3(a) from the analytical results obtained from Eq. (2) and (3) with the atomic model established from the previous simulations shown in Fig. 2. The same phenomenon can be observed from the numerical simulations (with microstructures included) shown in Fig. 3(b). Real parts (white dashed line in Fig. 3(a),(b)) of the eigenvalues of $\mathbf{H_{2}}$ agree well with the CPA positions in the spectra.  In both Fig. 3(a) and Fig. 3(b), the absorption loss ($1-|t+r|^2$ for symmetric incoming waves) of the hybridized system increases dramatically with increasing coupling (decreasing separation) below $\kappa_{PT}$. At $\kappa_{PT}$,  the presence of $\tilde{m}$ interferes with $\tilde{p}$ to the right size and gives $r=-\frac{1}{2}$ and $t=\frac{1}{2}$. This is the onset of CPA phase. Going beyond the transition point (larger coupling), the CPA peak splits into two. On the other hand, this splitting makes both atoms cannot be efficiently excited at $\omega_{0}$ and loss decreases at that frequency. In fact, the non-monotonic function of loss against coupling coefficient is a direct manifestation of spontaneous symmetry breaking, which is similar to the loss-induced optical transparency in a passive $\mathcal{PT}$-symmetric system deviated from the ideal $\mathcal{PT}$-symmetry\cite{R25}. Here, using metamaterials, we can actually establish an ideal $\mathcal{PT}$-symmetry in a completely passive system. CPA becomes possible besides the counterintuitive trend of loss function.

Besides $\mathbf{H_{2}}$ in revealing $\mathcal{PT}$-symmetry breaking process, $\mathbf{H_{1}}$ is also a potential candidate to display the same process. As predicted in Eq. (5) and Fig. 1, by taking into account the eigenvalue problem of $\mathbf{H_{1}}$, we need to fulfill the case $\gamma_{2}=-\gamma_{1}^{loss}=-1.75$ THz (the black vertical line in Fig. 1), \textit{i.e.} $\epsilon^{\prime\prime}=-0.156$, where potential amplifying media can be available in nature\cite{R34,R35,R36}. The emergence of effective  $\mathcal{PT}$-symmetry breaking process can be observed though transmission spectra, and  $t=0(r=-1)$ in spectra implies real frequency solution for eigenvalue of $\mathbf{H_{1}}$. It indicates the system without absorption loss, $A=1-|t+r|^2=0$, and transmission, $t=0$, with perfect reflection with and a $\pi$-phase shift, $r=-1$. The phase transition point in this configuration is at $\kappa_{\mathcal{PT}}=1.75$ THz ($h=98.4$ nm). Illustrated in Fig. 3(c), and numerically verified in Fig. 3(d),  the phase transition point is at $h=97.5$nm, $t=0$ ($r=-1$) in the transmission spectra agrees well with the real parts of the eigenvalues of $\mathbf{H_{1}}$. Below the phase transition point ($\kappa<\kappa_{\mathcal{PT}}$), the eigenvalue of $H_{1}$ implies no cross point in the real frequency axis, a complex conjugate pair, so transmission is dominated by the bright atom dipole oscillation, although $t\approx0$, the eigenvalue of $\mathbf{S}$ is not $-1$ (Absorption loss is not zero). Above the phase transition point $\kappa_{\mathcal{PT}}$, the eigenvalue of $H_{1}$ implies two real eigenfrequencies, so $t=0$ ($r=-1$) can be observed at two different frequencies in the transmission spectra.
Real parts (white dashed line in Fig. 3(c),(d)) of the eigenvalues of $\mathbf{H_{1}}$  match the $t=0$ positions well in spectra. The effective spontaneous $\mathcal{PT}$-symmetry breaking for $\mathbf{H_{1}}$ can thus be observed in the spectra. It is worth to note that the transmission exhibits maximum value at the position $\kappa_{s}=\sqrt{-\gamma_{2}\gamma_{1}}=5.12$ THz ($h=73.8$ nm for analytical prediction and $h=72.5$ nm for numerical results) at $\omega_{0}$, because gain behavior of the system appears, when pass over the phase transition point $\kappa_{\mathcal{PT}}$. As we increase $\kappa$, we can encounter a real eigenvalue of $H_{0}$, indicating pole of $\mathbf{S}$, which implies the system shows laser action at this parameter position\cite{R30,R34,R35,R36}, as indicted in the cross point between the black vertical line indicating $\mathcal{PT}$-symmetric $H_{1}$ and the blue
curve displaying real eigenvalue of $H_{0}$ in Fig. 1. The laser action results from non-ideal $\mathcal{PT}$-symmetry of $H_{0}$ whose origin will be discussed in the following.

For $\mathbf{H_{0}}$, ideal $\mathcal{PT}$-symmetry condition $\gamma_{2}=-\gamma_{1}$ implies that huge gain is required to compensate metal losses. This is difficult to be satisfied using natural materials. However, in a practical (non-ideal) system, even though ideal condition is not satisfied, the eigenvalues still split in the same way as an ideal $\mathcal{PT}$-symmetric Hamiltonian. Only an additional constant bias in the imaginary part of the eigenvalues occurs. Similar strategies have been employed in studying non-indeal $\mathcal{PT}$-symmetry in passive systems\cite{R25,R27}. Then, phase transition (branching) point is not on the real frequency axis, and is shifted into the lower-half of the complex $\omega$-plane. One of the eigenvalues (the one with increasing imaginary part) will cross the real frequency axis, corresponding to the original optical phenomenon, \textit{e.g.} laser action for the case of $\mathbf{H_{0}}$ as a pole of $\mathbf{S}$ at $\kappa_{s}=\sqrt{-\gamma_{1}\gamma_{2}}$\cite{R30,R34,R35,R36}. It is the solid blue curve in Fig. 1 and is lying below branching condition (branching occurs when $\kappa>\kappa_{\mathcal{PT}}$) where $\kappa_{\mathcal{PT}}=(\gamma_{1}-\gamma_{2})/2$(dashed green line). In contrary, ideal $\mathcal{PT}$-symmetric $\mathbf{H_{0}}$ has the poles of $\mathbf{S}$ on the real frequency axis beyond branching condition (vertical solid blue line) at two different frequencies. As an example, $\gamma_{2}=-0.5$ THz ($\epsilon^{\prime\prime}=-0.1$), Fig. 4 gives the outgoing intensity of the system with symmetric incoming light as a function of $\kappa(h)$. Laser action collapses to one position at $\omega_{0}$, $h=89.4$ nm for theoretical prediction and $h=90.0$ nm for numerical results. The branching phenomenon still occurs (without real poles of $\mathbf{S}$) and agrees well with the real parts of the eigenvalues of $\mathbf{H_{0}}$ (black dashed lines), as displayed in Fig. 4. Laser action and branch phenomenon is the direct demonstration of non-ideal $\mathcal{PT}$-symmetry for $\mathbf{H_{0}}$. Besides $\mathbf{H_{0}}$, non-ideal $\mathcal{PT}$-symmetry can also occur for $\mathbf{H_{1}}$ and $\mathbf{H_{2}}$. In Fig. 1, the real eigenvalue of  $\mathbf{H_{2}}$ (zero of $\mathbf{S}$), \textit{i.e.} CPA, is at $\kappa_{o}=\sqrt{\gamma_{2}(\gamma_{1}-2\gamma_{1}^{loss})}$ (solid red curved line) while the branching condition is at $\kappa_{\mathcal{PT}}= (\gamma_{1}+\gamma_{2}-\gamma_{1}^{loss})/2$ (dashed purple line). For $\mathbf{H_{1}}$, perfect reflection, $r=-1,t=0$ (one $\mathbf{S}$ eigenvalue is $-1$) is at $\kappa_{-1}=\sqrt{-\gamma_{2}\gamma_{1}^{loss}}$ (solid black curved line) while the branching condition is at $\kappa_{\mathcal{PT}}=(\gamma_{1}^{loss}-\gamma_{2})/2$ (dashed black line).
So, for a non-ideal $\mathcal{PT}$-symmetric system, the optical phenomenon (\textit{e.g.} laser, CPA or $\pi$-reflector) associated to $\mathbf{H_{n}}$ collapses to a single point (with a $\kappa$ a bit less than the one triggers branching) instead of the two branches.

\begin{figure}
\centerline{\includegraphics[width=8.0 cm]{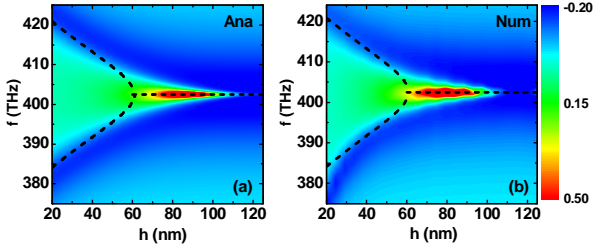}} \caption{(Color
online) Output intensity (color map) of the metamaterial $log_{10}|t+r|^2$ revealing spontaneous symmetry breaking of non-ideal $\mathcal{PT}$-symmetry for $\mathbf{H_{0}}$ and the real part of the eigenfrequencies for $\mathbf{H_{0}}$(dashed lines). (a) Analytical model; (b)Numerical simulations with metamaterial microstructures.}
\end{figure}

In the above discussion, $\kappa$ is the external parameter in demonstrating  $\mathcal{PT}$-symmetry. If we change $\gamma_{2}$ with a fixed $\kappa$ instead (travelling in a horizontal direction in Fig.1), the metamaterial can function as a laser, the cross point for $\mathbf{H_{0}}$, or a CPA, the cross point for $\mathbf{H_{2}}$ at $\omega_{0}$. As an example for demonstration, $h=75$ nm, besides appearance of laser action and CPA in spectra, schematically shown in Fig. 5, above CPA position, an increase of loss (increasing $\gamma_{2}$) leads to more output of the system, which is similar to loss induced optical transparency in a non-ideal passive $\mathcal{PT}$ system\cite{R25}. Similar counter-intuitive trend also occurs beyond the laser position, an increase in gain (decreasing negative $\gamma_{2}$) causes less output observed in the spectra.
\begin{figure}
{\includegraphics[width=6.0 cm]{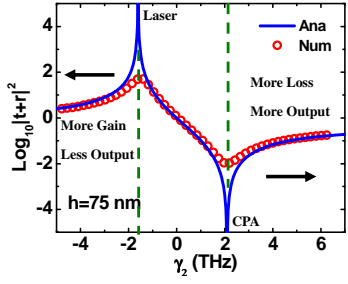}} \caption{(Color
online) Analytical (solid line) and numerical (dashed line) outgoing spectra, as a function of $\gamma_{2}$ for the metamaterial at $\omega_{0}$. }
\end{figure}

In conclusion, we investigated the effective $\mathcal{PT}$-symmetric Hamiltonians using a metamaterial system to reveal spontaneous symmetry breaking processes. Theoretical and numerical results clearly confirm the analog of spontaneous breaking of $\mathcal{PT}$-symmetry using metamaterials. Our results not only open up a new perspective towards testing the intriguing properties of spontaneous $\mathcal{PT}$-symmetry breaking in non-Hermitian quantum systems, but also provide feasibility for future applications in plasmonic and metamaterial systems, such as laser action\cite{R30,R34,R35,R36},CPA with sub-wavelength thickness\cite{R32},and asymmetric transmission\cite{R37}. Moreover, by using metamaterials with both electric and magnetic responses, it becomes possible to establish an ideal $\mathcal{PT}$-symmetry by exploring the balance between scattering and absorption loss using a passive system.


\appendix*
\section{Appendix}
\noindent The metamaterial in this work is modeled as a planar array of electric atoms coupled with an array of magnetic atoms on the same plane with surface normal vector $\hat{z}$ and a unit cell of area $A$. For simplicity, we assume the plane is excited by normally incident plane waves with electric field in the x-direction ($\mathcal{E_\text{x}}$) and magnetic field in the y-direction ($\mathcal{H_\text{y}}$). Suppose each electric atom only generates electric dipole moment ($p$) along the $x$-direction dominated by a single resonance while each magnetic atom only generates magnetic dipole moment ($m$) along its principal direction $\hat{y}\cos(\alpha)+\hat{z}\sin(\alpha)$ also dominated by a single resonance (at a general angle $\alpha$ with $\pi/2$ being the special case of a completely dark magnetic atom). Then, the response of these atoms can be described using a simple dipolar model as
\begin{equation}
\left(\begin{array}{cc}
    \omega _1 - \omega - i \gamma _1 & \kappa\\
    \kappa & \omega _2 - \omega - i \gamma _2 \\
  \end{array}
\right)\left(\begin{array}{cc}
   \tilde{p}  \\
    i\tilde{m} \\
  \end{array}
\right)=\left(\begin{array}{cc}
   \sqrt{a_1}\mathcal{E}_x \\
    i\sqrt{a_2}\mathcal{H} \\
  \end{array}
\right),
\end{equation}
where $\tilde{p}=k p/(2A\sqrt{a_{1}})$ and $\tilde{m}=k m/(2A\sqrt{a_{2}})$ with $k$ being the wavenumber in vacuum and the positive numbers $\omega_i$, $a_i$, $\gamma_i$ being the resonance frequencies, oscillator strength and total resonance linewidth of the $i$-th atom ($i=1$ or $2$ for electric or magnetic atom). $\kappa$ is the coupling parameter (controlled by the displacement between the electric and the magnetic atom in a unit cell), is assumed as a real number in the long wavelength limit in this work.
We have employed the Heaviside-Lorentz unit system and the normalization ($\tilde{p}$ and $\tilde{m}$) is chosen so that $\omega_i$, $a_i$, $\gamma_i$ and $\kappa$ are in the same unit of frequency.

If we use $a_{\pm}$ to indicate the E-field amplitudes of the incoming plane waves in the positive and negative  $z$-direction, the incident fields ($\mathcal{E_\text{x}}$ and $\mathcal{H}$)  applying on the electric and magnetic atoms can be expressed as
\begin{subequations}
\begin{equation}
\mathcal{E_\text{x}}=a_{+}+a_{-},
\end{equation}
\begin{equation}
\mathcal{H}=\mathcal{H_\text{y}}\cos{\alpha}=(a_{+}-a_{-})\cos{\alpha}.
\end{equation}
\end{subequations}

\noindent Substituting A.7 into A.6 gives the atomic response to the incoming plane-wave amplitudes as
\begin{equation}
\left(\begin{array}{cc}
    \omega _1 - \omega - i \gamma _1 & \kappa\\
    \kappa & \omega _2 - \omega - i \gamma _2 \\
  \end{array}
\right)\left(\begin{array}{cc}
   \tilde{p}  \\
    i\tilde{m} \\
  \end{array}
\right)=\left(\begin{array}{cc}
   \sqrt{a_1}(a_{+}+a_{-}) \\
    i\sqrt{a'_2}(a_{+}-a_{-}) \\
  \end{array}
\right),
\end{equation}
\noindent where $a'_2$ is a postive number defined as $a'_2=a_2 \cos^2{\alpha}$.

On the other hand, the scattering plane waves (with  $s_{\pm}$ being the E-field amplitudes in the positive and negative $z$-direction) are the radiation plane waves generated by the dipole moments:

\begin{subequations}
\begin{equation}
s_{+}+s_{-}=2i \sqrt{a_1}\tilde{p},
\end{equation}
\begin{equation}
s_{+}-s_{-}=2i \sqrt{a'_2}\tilde{m}.
\end{equation}
\end{subequations}
\noindent If we use $b_{\pm}=a_{\pm}+s_{\pm}$ to indicate the E-field amplitudes of the outgoing plane waves in the positive and negative $z$-direction, substituting A.9 and A.8 into it expresses the outgoing plane waves in terms of the dipolar moments as

\begin{equation}
\begin{aligned}
&\left(\begin{array}{cc}
    \omega _1 - \omega - i \gamma _1 + 2i a_1 & \kappa\\
    \kappa & \omega _2 - \omega - i \gamma _2 +2i a'_2\\
  \end{array}
\right)\left(\begin{array}{cc}
   \tilde{p}  \\
    i\tilde{m} \\
  \end{array}
\right) \\
&=\left(\begin{array}{cc}
   \sqrt{a_1}(b_{+}+b_{-}) \\
    i\sqrt{a'_2}(b_{+}-b_{-}) \\
  \end{array}
\right).
\end{aligned}
\end{equation}

In fact, we can also express the atomic response using the incoming and scattering plane waves by substituting A.9 into A.8 as
\begin{equation}
\frac{1}{2}\left(\begin{array}{cc}
    s_{+}+s_{-} \\
    s_{+}-s_{-} \\
  \end{array}
\right)=D\left(\begin{array}{cc}
   a_{+}+a_{-}  \\
    a_{+}-a_{-}\\
  \end{array}
\right)
\end{equation}

\noindent where the matrix $D$ is defined by

\begin{equation}
\begin{aligned}
& D=i
\left(\begin{array}{cc}
    \sqrt{a_{1}} & 0 \\
    0 & \sqrt{a'_{2}} \\
  \end{array}
\right) \times \\
& \left(\begin{array}{cc}
    \omega _1 - \omega - i \gamma _1 & i\kappa\\
    -i\kappa & \omega _2 - \omega - i \gamma _2 \\
  \end{array}
\right)^{-1}
\left(\begin{array}{cc}
    \sqrt{a_{1}} & 0 \\
    0 & \sqrt{a'_{2}} \\
  \end{array}
\right)
\end{aligned}
\end{equation}

\noindent If the system is without gain, we have the passivity condition requiring $\mid a_{+}\mid^{2}+\mid a_{-}\mid^{2}\geq\mid b_{+}\mid^{2}+\mid b_{-}\mid^{2}$.
By substituting A.11 into it for arbitrary excitaiton, after some algebra, the passivity condition means
$D^{-1}+(D^{-1})^{\dag}+2I$ being a negative-definite matrix. Equivalently, it means $\gamma_1>a_1$ and $\gamma_2>a'_2$ directly obtained from A.12. Therefore, for the general system, we can decompose the total resonance linewidth by $\gamma_{i}=\gamma_{i}^{s}+\gamma_{i}^{loss} (i=1,2)$ where the part due to scattering is
$\gamma_{1}^{s}=a_{1}$, $\gamma_{2}^{s}=a'_{2}$
and $\gamma_{i}^{loss}$ is the part due to absorption loss when it is positive.
With such decompositions of the linewidths, A.8 becomes Eq. 2 and A.10 becomes Eq. 3 in text immediately.

\end{document}